\documentclass[journal]{IEEEtran}
\usepackage{cite,graphicx,amssymb,amsmath,color,textcomp,tabularx}
\usepackage{multicol}
\usepackage{graphicx}
\usepackage{epstopdf}
\usepackage{stfloats}
\usepackage{algorithm,algorithmic}
\newtheorem{theorem}{Theorem}

\newtheorem{proposition}{Proposition}
\IEEEoverridecommandlockouts

\begin{document}
\title{\Huge{Secrecy Sum Rate Maximization in Non-Orthogonal Multiple Access}}
\author{Yi Zhang, Hui-Ming Wang, \IEEEmembership{Senior Member,~IEEE}, Qian Yang, and Zhiguo Ding, \IEEEmembership{Senior Member,~IEEE}\vspace{-8pt}
\thanks{
Manuscript received December 15, 2015; accepted February 27, 2016. The associate editor coordinating the review of this paper and approving it for publication was K. Tourki.

The work
was supported by
the Fund for the Author of National Excellent Doctoral Dissertation of China under Grant 201340,
the New Century Excellent Talents Support Fund of China under Grant NCET-13-0458 and by the Young Talent Support Fund of Science and Technology of Shanxi Province under Grant 2015KJXX-01.
The work of Z. Ding was supported by the UK EPSRC under Grant No. EP/L025272/1 and by H2020-MSCA-RISE-2015 under Grant No. 690750.
\emph{(Corresponding author: Hui-Ming Wang.)}

Y. Zhang, H.-M. Wang and Q. Yang are with School of Electronic and Information Engineering, Xi'an Jiaotong University, Xi'an 710049, China (e-mail: yi.zhang.cn@outlook.com; xjbswhm@gmail.com; qian-yang@outlook.com).

Z. Ding is with the Department of Electrical Engineering, Princeton University, Princeton, NJ 08544, USA, and also with the School of Computing and Communications, Lancaster University, LA1 4YW, U.K (e-mail: z.ding@lancaster.ac.uk).
}\vspace{-8pt}}
\maketitle

\begin{abstract}
Non-orthogonal multiple access (NOMA) has been recognized as a promising technique for providing high data rates in 5G systems.
This letter is to study physical layer security in a single-input single-output (SISO) NOMA system consisting of a transmitter, multiple legitimate users and an eavesdropper.
The aim of this letter is to maximize the secrecy sum rate (SSR) of the NOMA system subject to the users' quality of service (QoS) requirements. We firstly identify the feasible region of the transmit power for satisfying all users' QoS requirements. Then we derive the closed-form expression of an optimal power allocation policy that maximizes the SSR. Numerical results are provided to show a significant SSR improvement by NOMA compared with conventional orthogonal multiple access (OMA).
\end{abstract}
\begin{IEEEkeywords}
Non-orthogonal multiple access, physical layer security, power allocation, optimization.
\end{IEEEkeywords}

\section{Introduction}
NOMA has been recognized as a promising band efficient candidate for 5G wireless systems \cite{Noma_Performace}. Different from conventional orthogonal multiple access (OMA) such as time-division multiple access (TDMA), NOMA exploits the power domain to serve multiple users simultaneously. Early work on NOMA has mainly focused on the enhancement of spectral efficiency. In\cite{Cooeprative_NOMA,CTS}, different cooperative NOMA schemes have been investigated, respectively, for improving transmission reliability. In\cite{CRS}, resource allocation has been considered in a relaying system using the NOMA principle. In\cite{MIMONOMA}, multiple-antenna technologies have been used to further improve the performance of NOMA systems.

Due to the broadcast nature of wireless communications, information exchange between transceivers is vulnerable to eavesdropping, which poses a challenge to realize secure wireless transmission. Recently, physical layer security, which achieves secure transmissions by exploiting the dynamics in the physical layer, has drawn considerable attention\cite{Wyner, zongsu}. Naturally, this new concept of security can be applied to NOMA in order to realize robust secure transmission.

In this letter, we investigate the maximization of the secrecy sum rate (SSR) of a SISO NOMA system, where each user has a predefined quality of service (QoS) requirement.
To the best of our knowledge, this is the first effort to study physical layer security in NOMA. We firstly identify the feasible region of the transmit power for satisfying all users' QoS requirements. Then we derive the closed-form expression of an optimal power allocation policy that maximizes the SSR. Numerical results are provided to show a significant SSR improvement by NOMA compared with conventional orthogonal multiple access (OMA).

\section{System Model}
Consider a downlink system consisting of a transmitter, $M$ legitimate users and a passive eavesdropper. Each node in the system is equipped with a single antenna. The channel gain from the transmitter to the $m$-th user is denoted by $h_m=d_m^{-\frac{\alpha}{2}}g_m$, where $g_m$ is the Rayleigh fading channel gain, $d_m$ is the distance between the $m$-th user and the transmitter, and $\alpha$ is the path loss exponent. Similarly, the channel gain from the transmitter to the eavesdropper is modeled as $h_e=d_e^{-\frac{\alpha}{2}}g_e$.
The instantaneous channel state information (CSI) of each user is known at the transmitter while that of the eavesdropper is unknown. Without loss of generality, the channel gains can be sorted as $0<\left|h_1\right|^2\leq\left|h_2\right|^2\leq...\left|h_{M_e}\right|^2\leq\left|h_e\right|^2<\left|h_{M_e+1}\right|^2...\leq\left|h_M\right|^2$,
where $M_e$ denotes the number of the legitimate users whose channel gains are not greater than that of the eavesdropper.
It should be pointed out that the transmitter neither knows $\left|h_e\right|^2$ nor $M_e$.

Employing the NOMA scheme\cite{Noma_Performace,NOMA_SISO_DING}, the transmitter broadcasts a linear combination of $M$ signals to its users. The transmitted superposition signal can be expressed as $\sum_{m=1}^{M}\sqrt{\gamma_mP}s_m$, where $s_m$ is the message for the $m$-th user, $P$ denotes the total transmit power, and $\gamma_m$ represents the power allocation coefficient, i.e., it is the ratio of the transmit power for the $m$-th user's signal to the total transmit power $P$. Each user has a predefined QoS requirement, which demands the transmitter to send the message to each user with a minimum data rate, respectively. Meanwhile, the eavesdropper tries to intercept all legitimate users' messages\footnote{The eavesdropper may be interested only in a specific user's message. But in this work, we make a more conservative assumption that the eavesdropper intercepts each user's message, because the transmitter neither knows which user the eavesdropper wants to wiretap nor the eavesdropper's CSI.}.

\subsection{Achievable Rates of Legitimate Users}
The users apply the successive interference cancellation (SIC) to decode their own signals: the $m$-th user will first detect the $i$-th user's message, $i<m$, and then eliminate this message from its observed mixture, in a successive way. The $i$-th user's message, $i>m$, will be treated as noise. Thus, the achievable rate of the $m$-th user, $1\leq m\leq M$, is given by \cite{NOMA_SISO_DING}
\begin{equation}\label{Rm}
    R_b^m=\log_2\left(1+\frac{P\left|h_m\right|^2\gamma_m}{P\left|h_m\right|^2\sum_{i=m+1}^M\gamma_i+\sigma_n^2}\right),
\end{equation}
where $\sigma_n^2$ is the power of additive noise.

\subsection{Secrecy Sum Rate of the NOMA System}
We denote $R^m_e$ as the achievable rate of the eavesdropper to detect the $m$-th user's message. Let $R^m_s$ and $R_s$ represent the secrecy rate of the $m$-th user and the SSR, respectively.
In this work, we make a pessimistic assumption that the first $m-1$ users' messages have already been decoded before the eavesdropper tries to decode the $m$-th user's message, which overestimates the eavesdropper's capability. Thereby, $R_s$ given below serves as a lower bound on the SSR correspondingly.
Then, $R^m_e$, $R^m_s$ and $R_s$ can be given by
\vspace{-4pt}
\begin{subequations}\label{Ree}
    \begin{align}
    &R^m_e=\log_2\left(1+\frac{P\left|h_e\right|^2\gamma_m}{P\left|h_e\right|^2\sum_{i=m+1}^M\gamma_i+\sigma_n^2}\right),\label{Re}\\
    &R^m_s = \left[R_b^m -R^m_e\right]^+,\\
    &R_s = \sum\nolimits_{m=1}^{M}R^m_s,\label{RsO}
    \end{align}
\end{subequations}
respectively, where $[\cdot]^+\triangleq \max(0,\cdot)$.
It can be verified that $R_b^m\leq R^m_e$ when $\left|h_m\right|^2\leq\left|h_e\right|^2$ \cite{NOMA_SISO_DING}, making $R^m_s$ be zero when $1\leq m\leq M_e$\footnote{Particularly, in the considered problem, the secure transmission for the $m$-th user, $1\leq m\leq M_e$, should be guaranteed by using conventional cryptography implemented at upper layers.}.
Then, $R_s$ can be rewritten as
\begin{equation}\label{Rs}
    R_s = \sum\nolimits_{m=M_e+1}^{M}\left(R_b^m -R^m_e\right).
\end{equation}

\section{Secrecy Sum Rate Maximization}
In this section, we propose a power allocation policy to maximize the SSR of the system subject to all users' QoS requirements. Particularly, the closed-form expressions for the optimal power allocation coefficients $\{\gamma_m^{\textrm{Opt}}\}_{m=1}^M$, which maximize the SSR, are derived analytically.

We denote $Q_m$ as the minimum data rate required by the $m$-th user and then the QoS constraints are characterized as
\begin{equation}\label{RmQm}
R_b^m\geq Q_m, ~~~1\leq m\leq M,
\end{equation}
which can be retransformed as
\begin{equation}\label{RmQm2}
\gamma_m\geq A_m\left(P\left|h_{m}\right|^2\sum\nolimits_{i=m+1}^{M}\gamma_i+\sigma_n^2\right),~1\leq m\leq M,
\end{equation}
where $A_m\triangleq\frac{2^{Q_m}-1}{P\left|h_{m}\right|^2}$. Then, the SSR maximization problem can be formulated as
\vspace{-6pt}
\begin{subequations}\label{SEE_MV_Original}
    \begin{align}
    &~~~~~~~~~\max_{\gamma_m,1\leq m\leq M}~R_s\label{SEE_First} \\
    &\textrm{s.t.}~~\sum\nolimits_{m=1}^{M}\gamma_m\leq 1~~\textrm{and}~~(\ref{RmQm2})\label{gammaequ1}.
    \end{align}
\end{subequations}

Due to the QoS requirements, there must exist a minimum transmit power, denoted by $P_{\textrm{min}}$, that satisfies all users' QoS requirements. Then problem (\ref{SEE_MV_Original}) is feasible only when $P\geq P_{\textrm{min}}$. As a result, it is important to identify the feasible region of the transmit power before dealing with problem (\ref{SEE_MV_Original}).

\vspace{-10pt}
\subsection{Minimum Transmit Power to Satisfy QoS Requirements}\label{sub1}
Let $P_m$ represent the power of the $m$-th user's signal, then the problem of seeking out $P_{\textrm{min}}$ is formulated as
\begin{subequations}\label{minp}
    \begin{align}
    &~~~~~~~~~~~~~~~P_{\textrm{min}}\triangleq\min_{P_m,1\leq m\leq M}~~\sum\nolimits_{m=1}^MP_m\label{Ob RmQm2} \\
    &\textrm{s.t.}~P_m\geq B_m\left(\left|h_{m}\right|^2\sum_{i=m+1}^{M}P_i+\sigma_n^2\right),1\leq m\leq M,\vspace{-36pt}\label{C RmQm2}
    \end{align}
\end{subequations}
where $B_m\triangleq\frac{2^{Q_m}-1}{\left|h_{m}\right|^2}$ and (\ref{C RmQm2}) is from the QoS constraints (\ref{RmQm}). 
\begin{theorem}\label{theo1}
The objective function in (\ref{Ob RmQm2}) is minimized when all the constraints in (\ref{C RmQm2}) are active.
\end{theorem}
\begin{IEEEproof}
     We prove this theorem by contradiction. Suppose that a set of transmit power of all users' signals $\{P^{*}_m\}_{m=1}^M$ is the optimal solution to problem (\ref{minp}) with at least one constraint in (\ref{C RmQm2}) inactive. Without loss of generality, we suppose that the $k$-th constraint in (\ref{C RmQm2}) is inactive, i.e.,
     \begin{equation}\label{proof_contra}
        P^*_k>B_k\left(\left|h_{k}\right|^2\sum\nolimits_{i=k+1}^{M}P^*_i+\sigma_n^2\right).
    \end{equation}
    Now, we create a new set $\{P^{**}_m\}_{m=1}^M$ by defining $P^{**}_m=P^*_m$ for $m\neq k$ while $P^{**}_k$ is set to the right-hand side (RHS) of (\ref{proof_contra}). Observing the structure of the constraints (\ref{C RmQm2}), we find out that the RHS of (\ref{C RmQm2}), for an arbitrary $m$, is a monotonically non-decreasing function of $P_i$ for $1\leq i\leq M$. As a result, the setting of $P^{**}_k$, whose value is less than $P^{*}_k$, ensures that all the constraints in (\ref{C RmQm2}) hold for the newly created set $\{P^{**}_m\}_{m=1}^M$. However, with the definition of $\{P^{**}_m\}_{m=1}^M$, we can obtain $\sum_{m=1}^{M}P^{**}_m<\sum_{m=1}^{M}P^{*}_m$, which contradicts to the assumption that $\{P^{*}_m\}_{m=1}^M$ is the optimal solution to problem (\ref{minp}). Thereby, we conclude that all the constraints in (\ref{C RmQm2}) must be active for minimizing (\ref{Ob RmQm2}) and thus the proof is complete.
\end{IEEEproof}

Note that when all the constraints in (\ref{C RmQm2}) are active, the optimal solution to problem (\ref{minp}), denoted by $\{P_m^{\textrm{Min}}\}_{m=1}^M$, can be calculated sequentially in the order $M,M-1,...,1$, since $P_m^{\textrm{Min}}$ is determined by $\{P_i^{\textrm{Min}}\}_{i=m+1}^M$ and $P_M^{\textrm{Min}} = B_M\sigma_n^2$ is a deterministic quantity. Then, we have $P_{\textrm{min}}=\sum\nolimits_{m=1}^MP_m^{\textrm{Min}}$ and the feasible region of the transmit power is $P\geq P_{\textrm{min}}$.

\vspace{-10pt}
\subsection{Optimal Power Allocation Policy}\label{sub2}
Having obtained $P_{\textrm{min}}$, we are going to deal with the SSR maximization problem in (\ref{SEE_MV_Original}) given $P\geq P_{\textrm{min}}$. By substituting (\ref{Rm}) and (\ref{Re}) into (\ref{Rs}), $R_s$ can be reformulated as follows.
\begin{align}
    R_{s}&=\log_2\left(P\left|h_{M_e+1}\right|^2\sum\nolimits_{i=M_e+1}^M\gamma_i+\sigma_n^2\right)\nonumber\\
    &-\log_2\left(P\left|h_{e}\right|^2\sum\nolimits_{i=M_e+1}^M\gamma_i+\sigma_n^2\right)\label{proof1_Rs}\\
    &+\sum\nolimits_{m=M_e+1}^{M-1}\left[\log_2\left(P\left|h_{m+1}\right|^2\sum\nolimits_{i=m+1}^M\gamma_i+\sigma_n^2\right)\right.\nonumber\\
    &~~~~~~~~~~~~~~~~\left.-\log_2\left(P\left|h_m\right|^2\sum\nolimits_{i=m+1}^M\gamma_i+\sigma_n^2\right)\right].\nonumber
\end{align}
For notational simplicity, we define
\begin{equation}
    \begin{aligned}
        &C_m\triangleq
        \begin{cases}
        P\left|h_{e}\right|^2, & m=M_e,\\
        P\left|h_{m}\right|^2, & M_e+1\leq m\leq M,
        \end{cases}\\
        &t_m\triangleq\sum\nolimits_{i=m+1}^M\gamma_i,~~M_e\leq m\leq M-1,\\
        &J_m(t_m)\triangleq\log_2\left(C_{m+1}t_m+\sigma_n^2\right)-\log_2\left(C_{m}t_m+\sigma_n^2\right).\nonumber
    \end{aligned}
\end{equation}
Then $R_{s}$ in (\ref{proof1_Rs}) can be recast as
\begin{equation}\label{proof1_Rs_simple}
    R_{s}=\sum\nolimits_{m=M_e}^{M-1}J_m(t_m).
\end{equation}

After the reformulation, we can see that problem (\ref{SEE_MV_Original}) has two important properties: 1) the objective function $R_s$ is the sum of $M-M_e$ non-convex subfunctions with a uniform expression; and 2) the arguments $\{\gamma_m\}_{m=1}^M$ are coupled with each other in the constraints (\ref{RmQm2}) in a complicated way.

In the following we propose an algorithm to solve problem (\ref{SEE_MV_Original}) by exploiting the properties of the optimization problem. Our basic idea is: we firstly solve the maximization problem of each subfunction $J_m(t_m)$ for $M_e\leq m\leq M-1$ individually subject to the constraints (\ref{gammaequ1}). Then we prove that the set of the optimal solutions to each maximization problem possesses a \emph{unique common solution}. In other words, we can find a unique solution that simultaneously maximizes $J_m(t_m)$ for each $m$ from $M_e$ to $M-1$ and satisfies all the constraints in (\ref{gammaequ1}), which consequently solves the problem.
More specifically, let $\Im_m$ represent the set of the optimal solutions that maximize $J_m(t_m)$ subject to all the constraints in (\ref{gammaequ1}), where $M_e\leq m\leq M-1$, then our goal is to prove $\Im_{M_e}\cap\Im_{M_e+1}\cap...\cap\Im_{M-1} = \{\{\gamma_m^{\textrm{Opt}}\}_{m=1}^M\}$, where $\{\gamma_m^{\textrm{Opt}}\}_{m=1}^M$ is the unique common solution of the $M-M_e$ optimization problems.

We now solve these optimization problems. We first transform the original problem of maximizing $J_m(t_m)$ by using its monotonicity. The first-order derivative of $J_m(t_m)$ is given by
\begin{equation}\label{lemma}
\frac{dJ_m(t_m)}{dt_m}=\frac{\left(C_{m+1}-C_{m}\right)\sigma_n^2}{\ln2\left(C_{m+1}t_m+\sigma_n^2\right)\left(C_{m}t_m+\sigma_n^2\right)}\geq0.
\end{equation}
This indicates that $J_m(t_m)$ is a monotonically increasing function of $t_m$. Thus, the maximization of $J_m$ is equivalent to the maximization of $t_m$. Then, the $M-M_e$ optimization problems mentioned above can be uniformly formulated as
\begin{subequations}\label{op_uniform}
    \begin{align}
        &\max_{\gamma_i, 1\leq i\leq M}~~t_m\\
    \textrm{s.t.}~&\gamma_i\geq A_i\left(P\left|h_{i}\right|^2\sum_{j=i+1}^{M}\gamma_j+\sigma_n^2\right),~1\leq i\leq M, \label{gamma_condition1}\\
    ~~&\sum\nolimits_{i=1}^{M}\gamma_i \leq1. \label{gamma_condition3}
    \end{align}
\end{subequations}
Problem (\ref{op_uniform}) is solved by the following proposition.
\begin{proposition}
The necessary and sufficient condition for the optimal solution to problem (\ref{op_uniform}) is that both the constraints in (\ref{gamma_condition1}) for $1\leq i\leq m$ and the constraint (\ref{gamma_condition3}) are active. The closed-form solution to problem (\ref{op_uniform}) is given by
\begin{subequations}\label{CFS_Prop}
    \begin{align}
    &\gamma_i=\frac{A_i\left[P\left|h_i\right|^2\left(1-\sum_{j=1}^{i-1}\gamma_j\right)+\sigma_n^2\right]}{2^{Q_i}},~ 1\leq i\leq m.\label{calculate gamma}\\
    &t_m=1-\sum\nolimits_{i=1}^{m}\gamma_i.\label{calculate tm}
    \end{align}
\end{subequations}
\end{proposition}
\begin{IEEEproof}
It is obvious that problem (\ref{op_uniform}) is convex and then the Karush-Kuhn-Tucker (KKT) conditions are necessary and sufficient for the optimal solution to problem (\ref{op_uniform}):
\begin{align}
    &\lambda =
    \begin{cases}
    \mu_k-\sum_{i=1}^{k-1}\mu_iA_iP\left|h_{i}\right|^2, & 1\leq k \leq m,\\
    \mu_k-\sum_{i=1}^{k-1}\mu_iA_iP\left|h_{i}\right|^2+1, & m< k\leq M,
    \end{cases}\label{KKT_lambda}\\
&\mu_i\left[A_i\left(P\left|h_{i}\right|^2\sum\nolimits_{j=i+1}^{M}\gamma_j+\sigma_n^2\right)-\gamma_i\right]=0,~1\leq i\leq M,\\
&\mu_i\geq 0,~~~~~1\leq i\leq M, \\
&\lambda\left(\sum\nolimits_{i=1}^{M}\gamma_i-1\right)=0,\\
&\lambda\geq 0, \label{lambdan0}
\end{align}
where $\{\mu_i\}_{i=1}^M$ and $\lambda$ are the Lagrange multipliers for the inequality constraints (\ref{gamma_condition1}) and (\ref{gamma_condition3}), respectively. In order to prove that the constraints in (\ref{gamma_condition1}) are active for $1\leq i\leq m$ and that the constraint (\ref{gamma_condition3}) is active, it is necessary and sufficient to prove $\mu_i\neq 0$ for $1\leq i\leq m$ and $\lambda\neq0$, respectively. To this end, we firstly demonstrate $\mu_1\neq0$ by contradiction:

Supposing $\mu_1=0$ and setting $k=1$ in (\ref{KKT_lambda}), we obtain
\begin{equation}\label{v_lambda}
    \lambda=\mu_1=0.
\end{equation}
Substituting (\ref{v_lambda}) into (\ref{KKT_lambda}) yields
\begin{equation}\label{v_mu}
\mu_k=\sum\nolimits_{i=1}^{k-1}\mu_iA_iP\left|h_{i}\right|^2,~~~~1\leq k\leq m.
\end{equation}
Obviously, (\ref{v_mu}) implies that $\mu_k=0$ for $1\leq k\leq m$, because $\mu_1=0$ and $\mu_k$ can be calculated sequentially in the order $2,3,...,k$. However, under the condition that $\mu_k=0$, $1\leq k\leq m$, we set $k=m+1$ in (\ref{KKT_lambda}) and then obtain $\lambda=\mu_{m+1}+1>0$, which contradicts to (\ref{v_lambda}) obtained using the assumption $\mu_1=0$. Thus, we conclude that $\mu_1\neq0$ and
\begin{equation}\label{v_t_lambda}
\lambda=\mu_1\neq0,
\end{equation}
which indicates the inequality constraint (\ref{gamma_condition3}) must be active.

We then substitute (\ref{v_t_lambda}) into (\ref{KKT_lambda}) for $1\leq k\leq m$ and obtain
\begin{equation}
\mu_k=\sum\nolimits_{i=1}^{k-1}\mu_iA_iP\left|h_{i}\right|^2 + \lambda,~~~~1\leq k\leq m,
\end{equation}
which demonstrates that $\mu_k>0$ for $1\leq k\leq m$, since $\lambda$ is a positive number according to (\ref{lambdan0}) and (\ref{v_t_lambda}). Thereby, the constraints in (\ref{gamma_condition1}) are active when $1\leq k\leq m$.

In order to obtain the closed-form solution of problem (\ref{op_uniform}), we replace $\sum_{j=i+1}^{M}\gamma_j$ with $1-\sum_{j=1}^{i}\gamma_j$ in (\ref{gamma_condition1}) since (\ref{gamma_condition3}) is proved to be active. Setting the constraints in (\ref{gamma_condition1}) active for $1\leq i\leq m$, we have the expressions of $\{\gamma_i\}_{i=1}^m$ and of $t_m$ given by (\ref{calculate gamma}) and (\ref{calculate tm}), respectively. According to (\ref{calculate gamma}), $\gamma_i$ can be obtained by recursive calculation in the order $1,2,...,m$. Thereby, the closed-form expression for the optimal solution, which maximizes $t_m$, can be obtained when both the constraints in (\ref{gamma_condition1}) for $1\leq i\leq m$ and the constraint (\ref{gamma_condition3}) are active. The proof is complete.
\end{IEEEproof}

Based on Proposition 1, which solves problem (\ref{op_uniform}) with the closed-form solution given by (\ref{CFS_Prop}), the following theorem gives the unique solution for maximizing $R_s$.
\begin{theorem}
The unique optimal power allocation coefficients $\{\gamma_m^{\textrm{Opt}}\}_{m=1}^M$, which maximize $R_s$, are given by
    \begin{equation} \label{gammaMaxS}
        \gamma_m^{\textrm{Opt}}=
        \begin{cases}
        \frac{A_m\left[P\left|h_m\right|^2\left(1-\sum_{i=1}^{m-1}\gamma_i^{\textrm{Opt}}\right)+\sigma_n^2\right]}{2^{Q_m}},
        &1\leq m< M,\vspace{4pt}\\
        1 - \sum_{i=1}^{M-1}\gamma_i^{\textrm{Opt}},&m=M.
        \end{cases}
    \end{equation}
\end{theorem}
\begin{IEEEproof}
According to Proposition 1, when both the constraints in (\ref{gamma_condition1}) for $1\leq i\leq m$ and the constraint (\ref{gamma_condition3}) are active, the arguments $\{\gamma_i\}_{i=1}^m$ are uniquely determined with the formulas (\ref{calculate gamma}). This implies that more power allocation coefficients are uniquely determined with the growth of $m$. In other words, the size of the set of the optimal solutions to problem (\ref{op_uniform}), i.e., $\Im_{m}$, becomes smaller as $m$ increases, which can be characterized as
\begin{subequations}
    \begin{align}
    &\mathfrak{}\Im_{M_e}\supset\Im_{M_e+1}\supset...\supset\Im_{M-1},\\
    &\Im_{M_e}\cap\Im_{M_e+1}\cap...\cap\Im_{M-1} = \Im_{M-1}.
    \end{align}
\end{subequations}

Accordingly, $\Im_{M-1}$ is the set of the optimal solutions that simultaneously maximize $J_m(t_m)$ for $M_e\leq m\leq M-1$, which consequently solves problem (\ref{SEE_MV_Original}). By setting $m=M-1$ in the closed-form solution provided by (\ref{CFS_Prop}) in Proposition 1, the first $M-1$ arguments $\{\gamma_i^{\textrm{Opt}}\}_{i=1}^{M-1}$ are uniquely determined with the formulas (\ref{calculate gamma}) in the order $1,2,...,M-1$ and the last argument $\gamma_M^{\textrm{Opt}}$ is uniquely determined with $\gamma_M^{\textrm{Opt}}=1-\sum_{i=1}^{M-1}\gamma_i^{\textrm{Opt}}$ since the constraint (\ref{gamma_condition3}) has been proved to be satisfied at equality. Thus the proof is complete.
\end{IEEEproof}

The analysis above demonstrates that with $P\geq P_{\textrm{min}}$, the optimal power allocation policy for maximizing the SSR of the NOMA system is to use the extra power $\left(P-P_{\textrm{min}}\right)$ only for increasing the $M$-th user's secrecy rate. This is because the $M$-th user has the best channel condition and thereby it is most likely to improve the SSR of the system. That is why the terms $h_e$ and $M_e$ do not appear in the closed-form solution given by (\ref{gammaMaxS}), namely, the proposed power allocation policy does not need the eavesdropper' CSI.

\section{Numerical Results}
In this section, we provide numerical results to demonstrate the SSR performance of the NOMA system by using the proposed power allocation policy. A TDMA system, where the time slots with equal duration are allocated to users individually, is used as a benchmark.
We randomly generate 50,000 channel realizations with parameters $g_m, g_e\sim\mathcal{CN}(0,1)$, $1\leq m\leq M$, $\alpha=3$, $d_m=d_e=80$m, $1\leq m\leq M$, and $\sigma_n^2=-70$ dBm. Particularly, when the value of the transmit power $P$ is not in the feasible region, the transmitter will not send the messages. In other words, when $P$ is not large enough to satisfy all users' QoS requirements, we set $R_s$ to zero.
\begin{figure}[!hbt]
    \vspace{-0.5em}
    \centering
        \includegraphics[height = 6.0cm, width = 7.2cm]{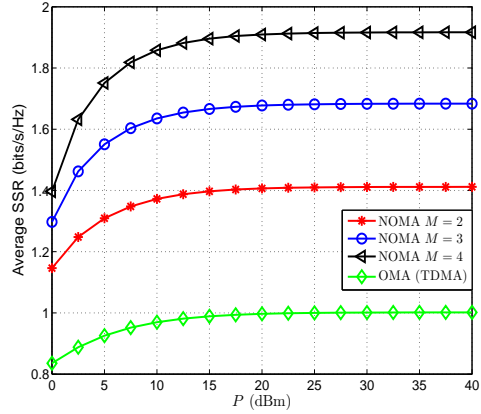}
    \vspace{-0.5em}
    \caption{Average SSR (bits/s/Hz) versus the transmit power $P$ for different numbers of users with parameters $Q_m=1$ bits/s/Hz, $1\leq m\leq M$.}
    \label{fig:SEE_P}
\end{figure}

Fig. \ref{fig:SEE_P} shows the average SSR versus the transmit power. One can see that NOMA outperforms conventional OMA and the performance gain becomes more significant as $M$ increases. This is because, a higher diversity gain is offered when $M$ is large, and higher spectral efficiency is achieved when more users are served simultaneously.

\begin{figure}[!hbt]
    \vspace{-0.5em}
    \centering
    \includegraphics[height = 6.0cm, width = 7.2cm]{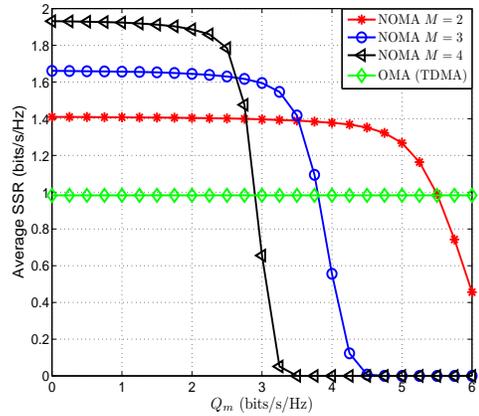}
    \vspace{-0.5em}
    \caption{Average SSR (bits/s/Hz) versus $Q_m$ (bits/s/Hz) for different numbers of users. The transmit power $P$ is set to 20 dBm.}
    \label{fig:SSR_QoS}
    \vspace{-0.5em}
\end{figure}

Fig. \ref{fig:SSR_QoS} shows the effect of the QoS requirements on the SSR performance. We can see that the SSR decreases as $Q_m$ increases. This is because, the increase of $Q_m$ requires the transmitter to use the extra power for improving the data rate of the users with poor channel conditions, which will obviously deteriorate the SSR performance. Further, as $Q_m$ becomes very large, the SSR approaches zero, since $P$ is not large enough to satisfy all users' QoS requirements and then the transmitter does not send the messages to the users.

\section{Conclusion}
In this letter, we studied physical layer security in a SISO NOMA system, in which each legitimate user has a predefined QoS requirement. We firstly identified the feasible region of the transmit power for satisfying all users' QoS requirements. Then, the optimal power allocation policy for maximizing the SSR was obtained in closed-form expressions. Numerical results showed that NOMA has the superior SSR performance compared with conventional OMA and the performance gain is more significant as the number of users increases. It is worth pointing out that the highly demanding QoS requirements will degrade the SSR performance of the system.
One promising future direction is to adopt multiple antenna settings to further enhance the security of NOMA systems.

\end{document}